# Physico-chemical study of polymer mixtures formed by a polycation and a zwitterionic copolymer in aqueous solution and upon adsorption onto negatively charged surfaces

**Laura Fernández-Peña [1,2], Eduardo Guzmán [1,3,*], Francisco Ortega [1,3], Lionel Bureau [4], Fabien Leonforte[5], Dandara Velasco [4], Ramón G. Rubio [1, 3,*] and Gustavo S. Luengo [5,*]**

[1] Departamento de Química Física, Facultad de Ciencias Química, Universidad Complutense de Madrid. Ciudad Universitaria s/n, 28040-Madrid (Spain)

[2] Centro de Espectroscopía y Correlación, Universidad Complutense de Madrid. Ciudad Universitaria s/n, 28040-Madrid (Spain)

[3] Instituto Pluridisciplinar, Universidad Complutense de Madrid. Paseo Juan XXIII 1, 28040-Madrid (Spain)

[4] Université Grenoble Alpes, CNRS, LIPhy. 38000-Grenoble (France)

[5] L'Oréal Research and Innovation, 93600-Aulnay-Sous-Bois (France)

\* Correspondence: eduardogs@quim.ucm.es-Tel.: +34 91 394 4107 (E.G.); rgrubio@quim.ucm.es-Tel: +34 91 394 4123 (R.G.R.); gluengo@rd.loreal.com-Tel: +33158317173 (G.S.L.)

**Abstract:**

The adsorption of mixtures of charged polymers onto solid surfaces presents a big interest in different technological and industrial fields, and in particular, in cosmetics. This requires to deepen on the most fundamental physico-chemical bases governing the deposition, which is generally correlated to the interactions occurring in solution. This work explores the interaction in solution of model polymer mixtures formed by a cationic homopolymer (poly(diallyl-dimethyl-ammonium chloride), PDADMAC) and a zwitterionic copolymer (copolymer of acrylic acid, 3-Trimethylammonium propyl methacrylamide chloride and acrylamide, Merquat 2003), and the adsorption of such mixtures onto negatively charged surfaces. The analysis of the interactions occurring in solution between both polymers performed using dynamic light scattering (DLS), electrophoretic mobility and viscosity measurements, combined with the study of the deposition of the layers of mixtures containing different weight fractions of each polymer using ellipsometry and quartz crystal microbalance with dissipation monitoring (QCM-D) has shown that the interpolymer complexes formed in solution, and their composition, governs the deposition onto the solid surface and the tribological properties of the adsorbed layers as shown the Surface Force Apparatus (SFA) experiments, allowing for a control of the physico-chemical properties and structure of the layers. Furthermore, the use of Self Consistent Mean Fields Calculations (SCF) confirms the picture obtained from the experimental studies of the adsorbed layers, providing a prediction of the distribution of the polymer chains within the adsorbed layers. It is expected that this study can help on the understanding of the correlations existing between the behavior of future associations of innovative and eco-sustainable polymers and their adsorption processes onto solid surface.

# 1. Introduction

Polymer adsorption onto solid surfaces impacts on many scientific and technological aspects (detergency, solubilisation, flotation, encapsulation or lubrication) relevant to different industries (e.g. personal care products, pharmaceuticals or oil industries). It has therefore stimulated an important research activity, both experimental and theoretical [1-12], in order to shed light on the physico-chemical bases underlying the interaction between polymers and surfaces, and their consequences on the physico-chemical properties and structure of the polymer-decorated surfaces [13-24]. However, most of the studies have dealt with the analysis of the adsorption process of single polymers [1-3] which presents a more limited interest from the technological and industrial perspectives because many consumer products involve very complex mixtures, including among other compounds several polymers, silicones, proteins, surfactants [13, 14, 25-27].

The adsorption of polyelectrolytes and their mixtures with oppositely charged and zwitterionic surfactants have been widely studied for more than 30 years. However, the adsorption of mixtures of polyelectrolytes has been less explored, even though it has been demonstrated that the use of a combination of several polymers can present significant benefits over a situation in which single polymers are used [28]. The adsorption of different mixtures of oppositely charged polymers onto solid surfaces has been studied in relation to their ability as flocculants for solid/liquid studies [29, 30]. These studies showed that the deposition of oppositely charged mixtures presents a better performance in flocculation than single polyelectrolytes [30, 31]. These processes rely on the deposition of inter-polyelectrolyte complexes particles, formed in solution as a result of the interaction of the oppositely charged polyelectrolytes, onto the surface of the colloidal particles. The adsorbed layers resulting from the deposition of the inter-polyelectrolyte complexes generally mirrors the composition of the complexes formed in solution [32]. This makes it necessary to explore the different physicochemical variables (nature of the ionic groups, polymerization degree, flexibility, charge density and concentration of the interacting polymers, molar ratio between the opposite charges or properties of the medium) affecting the formation of inter-polyelectrolyte complexes to exploit their ability for the modification of surfaces [31, 33].

The progressive substitution of charged molecules for neutral one or at least with zwitterionic character is interesting in order to reduce the level of eco-toxicity of consumer products. The use of zwitterionic materials, including zwitterionic polymers, appears as a very promising alternative for the modification of surfaces. This is because the deposition of zwitterionic polymers onto surfaces allows, in many cases, manufacturing smart materials with stimuli-responsiveness, and an antipolyelectrolyte-polyelectrolyte dual behavior depending on the solvent quality, temperature, pH or ionic strength [34-36]. Furthermore, it is known that zwitterionic polymers present a strong ability to self-coacervate in solution as a result of the inter- and intra-chain attractive electrostatic interactions and the entropy gain associated with the counterions release. Therefore, the formation of polymer aggregates reminiscent of the inter-polyelectrolyte complexes formed by oppositely charged polyelectrolytes can be expected in solutions containing zwitterionic polymers [37]. This suggests that having a zwitterionic character in the macromolecular structure can be a promising alternative when developing new bio-based and biodegradable cosmetic formulations following current market trends in eco-sustainability [38, 39]. New trends of industry focus to replace raw polymer materials derived from petroleum chemistry for new eco-friendly and biodegradable polymers, preferably from renewable, natural resources. However, this substitution requires maintaining (or improving) the performance of the products. Therefore, the understanding of the main physico-chemical bases underlying the performance of the currently used products becomes mandatory to accomplish successfully the challenge of manufacturing sustainable consumer products [40].

This work analyzes the effect of the concentration of the polymers on the adsorption of mixtures formed from a cationic homopolymer (poly(diallyl-dimethyl-ammonium chloride), PDADMAC) and a zwitterionic copolymer (copolymer of acrylic acid, 3-Trimethylammonium propyl methacrylamide chloride and acrylamide, Merquat 2003) onto negatively charged surfaces. We will focus on the impact on deposition of the possible association occurring between the two polymers in solution, i.e. possible formation of inter-polymer complexes. For this purpose, we will combine a set of measurements giving access to information about the behavior of the solutions with the analysis of the wet and dry thicknesses, and the water content of the adsorbed layers. Scheme 1 summarised the main aspects dealt in this work.

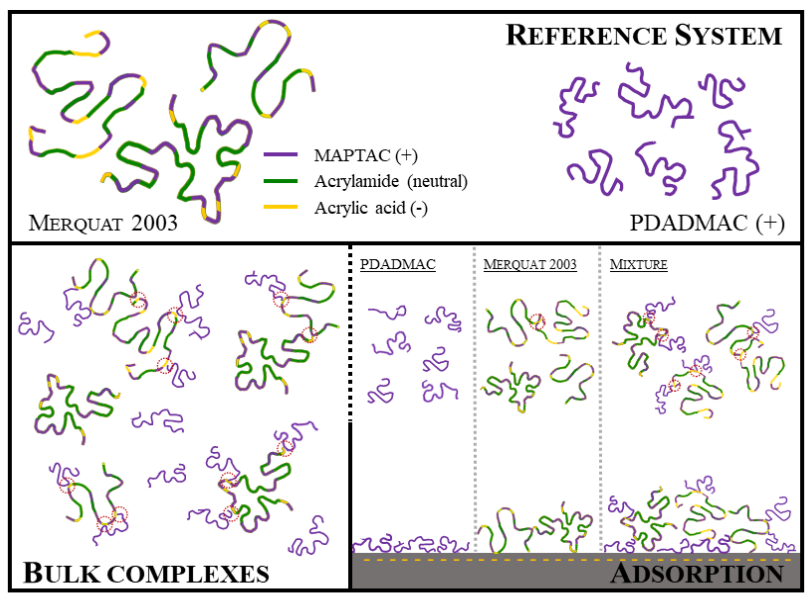

**Schme 1.** Sketch of the different aspects studied in this work.

## 2. Materials and Methods

### 2.1. Chemicals

Poly(diallyl-dimethyl-ammonium chloride), PDADMAC, with a molecular weight in the 100-200 kDa range, and the zwitterionic copolymer Merquat™ 2003 (copolymer of acrylic acid, 3-Trimethylammonium propyl methacrylamide chloride and acrylamide in molar ratio 10:40:50) with a molecular weight about 1200 kDa were purchased from Sigma-Aldrich (Saint Louis, MO, USA) and Lubrizol (Wickliffe, OH, USA), respectively. Polymers were used as received without any further purification. Figure 1 shows the molecular structures of both polymers.

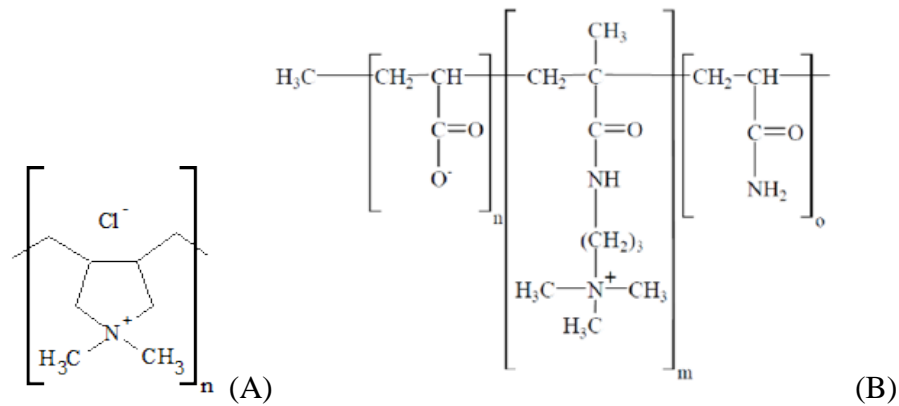

**Figure 1.** Molecular structure of the polymers studied in this work: PDADMAC (A) and Merquat™ 2003 (B).

Citric acid with a purity > 99.9% and NaOH, both purchased from Sigma-Aldrich (Saint Louis, MO, USA), were used for adjusting the pH of the solutions at 5.5, and NaCl (Sigma-Aldrich, Saint Louis, MO, USA, purity > 99.9%) was used for controlling the ionic strength of the solutions at 120 mM. It should be noted that pH and ionic strength were fixed in values similar to those of shampoos and closer to the values of the scalp [41].

Ultrapure deionized water used for cleaning and solution preparation was obtained by a multicartridge purification system AquaMAX$^{TM}$-Ultra 370 Series. (Young Lin Instrument Co., Ltd., Gyeonggi-do, South Korea), presenting a resistivity higher than 18 MΩ·cm, and a total organic content lower than 6 ppm.

*2.2. Methods*

2.2.1. Characterization of polymer solutions

Dynamic Light Scattering (DLS) measurements were performed using a Zetasizer Nano ZS (Malvern Instrument, Ltd., Malvern, United Kingdom). The DLS measurements were performed at $25^0$C using the red line of a He-Ne laser (wavelength, $\lambda = 632$ nm) in quasi-backscattering configuration (scattering angle, $\theta=173^0$). Before each measurement, the samples were filtered in a clean room using a 0.45 μm Nylon membrane (Millex®, Merck-Millipore, Burlington, MA, USA) to remove dust particles, and transferred to the measurement cells. DLS experiments rely on the determination of the normalized intensity auto-correlation function, $g^{(2)}(q,t)$, which can be represented for monodisperse scatterers with Brownian motion through an exponential-like decay [42]

$$g^{(2)}(q,t) = 1 + \beta e^{-2D^{app}q^2 t}, \quad (1)$$

with $\beta$ being an optical coherence factor which generally takes values close to 1, except for those cases in which the scattered intensity is low (generally due to a very low concentration, or the poor refractive index contrast between the scatterers and the solvent) and $D^{app}$ the apparent diffusion coefficient of the scatterers. $q = (4\pi n / \lambda)sin(\theta / 2)$ is the scattering wave-vector (with n being the refractive of the continuous phase). The apparent diffusion coefficient is correlated to the characteristic relaxation time of the scatterers $\tau$ as follows

$$D^{app}q^2 = 1/\tau. \quad (2)$$

It is worth mentioning that the above discussion considers the existence of only one relaxation mode in the dynamics of the solutions. However, additional contributions may appear, resulting in a more complex decay of the normalized intensity auto-correlation function than that described by Equation (1). This can be accounted for by the inclusion of additional exponential-like terms in the expression of the normalized intensity auto-correlation function, which requires a careful analysis of the experimental results on the basis of the CONTIN method [43].

For a spherical scatterers diffusing in a continuous Newtonian fluid, it is possible to correlate their apparent diffusion coefficient to their apparent hydrodynamic diameter using the Stoke-Einstein

$$d_h^{app} = \frac{k_B T}{3\pi \eta D^{app}}, \quad (3)$$

with $k_B$, $T$ and $\eta$ being the Boltzmann constant, the absolute temperature and the viscosity of the solvent respectively.

The evaluation of the viscosity of polymer solutions and its dependence with the concentration is very important for describing the solution flow properties, providing semi-quantitative information on the average size of individual polymer molecules in term of the hydrodynamic volume. This can be inferred from the intrinsic viscosity, $[\eta]$, of the solutions derived from the viscosity measurements obtained using a calibrated home-built capillary viscometer type Ubbelohde at 25°C [44, 45]. From the viscosity of the solutions, it can be obtained the intrinsic viscosity according to the Huggins empirical expression[46]

$$\eta_{red} = [\eta] + k_H[\eta]^2 c, \quad (4)$$

with $\eta_{red} = \eta_{sp}/c$ being the reduced viscosity, and $\eta_{sp} = (\eta/\eta_0)-1$ the specific viscosity. $k_H$ is the Huggins constant, and $\eta$ and $\eta_0$ are the apparent viscosities of the solution and the solvent obtained from the experimental measurements, respectively. The extrapolations to zero concentration are usually determined by plotting $\eta_{red}$ vs. $c$, which for solutions diluted enough results in a straight line

The effective charge density of the polymer chains in solution can be inferred from measurements of the electrophoretic mobility, $u_e$, by Laser Doppler velocimetry using a Zetasizer Nano ZS (Malvern Instrument, Ltd., Malvern, United Kingdom). The electrophoretic mobility is directly proportional to the zeta potential, $\zeta$, which gives a measurement of the effective charge of the polymer chains, by the Henry's equation [47]

$$u_e = \frac{2\varepsilon}{3\eta}\zeta f(\kappa a), \qquad (5)$$

where $\varepsilon$ represents the dielectric permittivity of the medium and $f(\kappa a)$ is the Henry function, which for particles big enough assumes a value of 1.5 (Smoluchowski equation) [48].

2.2.2. Characterization of the adsorbed layer

A dissipative quartz-crystal microbalance (QCM-D) from KSV (Model QCM Z-500, Espoo, Finlands) fitted with gold coated AT-cut quartz crystals (*Note: gold coated AT-cut quartz crystals were initially cleaned with piranha solution, 70% sulfuric acid/30% hydrogen peroxide, over thirty minutes, and then thoroughly rinsed with Milli-Q water*). The surface of the quartz sensor was modified by a self-assembled monolayer of 3-mercapto-propanesulfonic acid to obtain a negatively charged surface, which was then used for studying the adsorbed layers. QCM-D measures the impedance spectra of a quartz crystal for the fundamental frequency ($f_0$ = 5 MHz) and the odd overtones up to 11[th] (central frequency, $f_{11}$ = 55 MHz). The obtained impedance spectra were analyzed using a single layer model following the procedure described by Voinova et al. [49], which provides the effective acoustic thickness or hydrodynamic thickness of the adsorbed layer, $h_{ac}$. This procedure makes it possible to correlate the changes in the resonant frequency $\Delta f$ and dissipation factor $\Delta D$ of the different overtones with physical parameters of the layers (thickness $h_j$, density $\rho_j$, elasticity $\mu_j$ and viscosity $\eta_j$), according to the following two equations (*The fundamental frequency is not used for data analysis due to the noisy character of its signal*)

$$\Delta f = -\left(\frac{1}{2\pi\rho_0 h_0}\right)\left[\frac{\eta}{\delta} + h_j\rho_j\omega - 2h_j\left(\frac{\eta}{\delta}\right)\frac{\eta_j\omega}{\mu_j^2 + \omega^2\eta_j^2}\right] \quad (6)$$

and

$$\Delta D = -\left(\frac{1}{2\pi f\rho_0 h_0}\right)\left[\frac{\eta}{\delta} + 2h_j\left(\frac{\eta}{\delta}\right)^2\frac{\eta_j\omega}{\mu_j^2+\omega^2\eta_j^2}\right], \quad (7)$$

where $\rho_0$=2.65 g/cm$^3$ and $h_0$=14 mm represents the density of the quartz crystal and its thickness, respectively; $\eta$ and $\delta=(2\eta/\rho)^{0.5}$ are the viscosity of the bulk aqueous solution and the viscous penetration depth of the shear wave in the bulk liquid, respectively, with $\rho$=0.997 g/cm$^3$ being the density of the bulk liquid. $\omega=2\pi f$ is the angular frequency of the oscillation. Equations (6) and (7) were used for modelling the experimental data, using a Simplex algorithm to find the optimum solution. The $\rho_j$ value was fixed in an average value between polyelectrolyte and water densities. Thus, considering the density of the pure polyelectrolyte (around 1.2 g/cm$^3$ for PDADMAC according the supplier) and the water associated with polyelectrolyte layers (around 50-80% of the total weight of the film) [1-3], and the difficulties for obtaining the layer density independently, the choice of a fixed density for the layers around 1.1 g/cm$^3$ appears as a realistic approach (*Note: the use of $\rho_j$ values in the 0.75-1.25 g/cm$^3$ results in modification of the calculated thickness and viscoelastic moduli within the error bars*). $\mu_j$ and $\eta_j$ were obtained, together with the layer thickness, from the analysis of the experimental changes in frequency and dissipation obtained as a result of the layer adsorption. Assuming values in the ranges $10^3$-$10^6$ Pa and $10^{-3}$-0.1 Pa·s for $\mu_j$ and $\eta_j$, respectively.

An imaging null-ellipsometer from Nanofilm (Model EP3, Göttingen, Germany) was also used to determine the amount of material adsorbed onto the solid surfaces as the optical thickness, $h_{op}$. Ellipsometry experiments were carried out using a solid-liquid cell at a fixed angle of 60° using silica plates as substrate (Siltronix, Archamps, France). These substrates were treated with piranha solution for 30 minutes to create a charged surface similar to that of thiol-decorated gold surfaces [1]. The experimental variables measured in ellipsometry are the ellipsometric angles, $\Delta$ and $\Psi$, which are related to the ratio between the reflection coefficients for the parallel ($r_p$) and normal ($r_s$) components of the magnetic field derived by Fresnel, i.e. to the ellipticity $\rho^e$ [50, 51]

$$\rho^e = \frac{r_p}{r_p}e^{i\Delta}\tan\Psi, \quad (8)$$

The optical thickness and refractive index of the adsorbed layers are obtained from the experimental measurements by assuming a slab model describing the system. Here, a four-layer slab was used: the first layer corresponds to the silicon substrate having a refractive index n=4.1653-0.049i, the second layer is the native oxide layer with a refractive index n=1.4653, and a thickness obtained from the measurement of the bare silicon wafer in water. The outermost layer of the model (fourth layer) was the solution which was assumed to have a constant refractive index similar to that of the polymer solution (n~ 1.33) [22]. The third layer of the model corresponds to the adsorption layer. Once the slab is defined, the thickness and the refractive index are obtained as the pair of values that minimize the differences between the experimental values of the ellipsometric angles and those obtained solving the Fresnel's equation using the four-layer model [51, 52].

$h_{ac}$ and $h_{op}$ should not be considered as absolute thicknesses due to the heterogeneity of most of the polyelectrolyte layers, thus any discussion contained in this work considers $h_{ac}$ and $h_{op}$ as effective thicknesses that provide different information about the adsorbed amount within the layer [2, 5]. The combination of ellipsometry and QCM-D is important because of the different sensitivities of these techniques to the water. This is because whereas the QCM-D provides information on the total mass of the adsorbed layer, including both the polymer and the water associated with such layer, ellipsometry, which is based in the differences between the refractive indices of the layer and the medium, only gives information on the amount of adsorbed polymer. This difference leads to $h_{op} \leq h_{ac}$ and allows one to estimate the water content of the layers $x_w$ as [53, 54]

$$x_w = \frac{h_{ac} - h_{op}}{h_{ac}}. \qquad (9)$$

The frictional properties of the adsorbed films were evaluated by shear experiments using the home-built Surface Force Apparatus (SFA) described by Bureau [55]. To this aim, two freshly cleaved mica sheets, exhibiting negatively charged surfaces, were glued on cylindrical lenses of radius R=1 cm. Cylindrical lenses were then mounted into the SFA with their axis crossed at right angle (thus equivalent to a sphere/flat contact geometry), and polymers were left to adsorb from solution onto mica for 15min. After adsorption under quiescent conditions, the polymer solution was replaced by fresh solvent (containing 120 mM of NaCl), and three different types of experiments were performed: (i) Determination of the normal force vs. separation distance curve (velocity of approaching/retraction of piezo of 1 nm/s); (ii) evaluation of friction forces (under normal loads in the range 250-3000 μN, at a shear speed of 10 μm/s, and (iii) measurement of the velocity dependence of friction (in the velocity range 1-100 μm/s).

It should be noted that the adsorption of the here studied polymers onto charged surfaces can be characterized in terms of equilibrium thickness and water content, and such information can be obtained combining QCM-D and ellipsometry. However, an important drawback for combining both techniques is the existence of differences in the adsorption process onto surfaces with different physico-chemical properties. This issue is not a problem for the study of the adsorption of the polymers here considered because their adsorption onto QCM-D sensors functionalized with a charged thiol or coated with a silica layer was found to be rather independent on the chemical nature of the surface. This enables a direct comparison between the experiments performed in QCM-D and those performed in ellipsometry, as was discussed in our previous work [1]. In addition, many studies using the SFA [56-58] have shown that negatively charged mica surfaces can represent well the overall adsorption process taking place when polyelectrolytes are concerned, and hence it is possible to access to the lubrication properties of the layers [59, 60].

*2.3. Self Consistent Mean Field Calculations*

Self-Consistent-Field calculations method of Scheutjens and Fleer (SCF) [75-77] is known to be a powerful tool to rationalize self-assembly behavior of complex mixtures of polymeric-based materials. They are an efficient and accurate alternative to atomistic simulations for capturing relevant features of the thermodynamics and structures of such systems [78,79], and have been shown to be especially well suited to devise semi-quantitative predictions of the most probable structures of polyelectrolytes and polyelectrolyte-surfactant mixtures at equilibrium as ones in cosmetic industry, with potential interests in the bulk formulations as well as in adsorption onto biomimetic surfaces [7,24,25,80,81]. Based on those previous works, we intend to apply such a theoretically-informed numerical approach to rationalize the deposits of mixtures of PDADMAC:Merquat 2003.

SCF approach of Scheutjens and Fleer include chemical nature of the species where freely-jointed chains that mimic the polymer are considered. The interactions between species are encoded though the Flory-Huggins (FH) parameters, $\chi$, and the approach also accounts for the dielectric medium background, $\varepsilon$, and the valence of charges, $\nu$. The discretization scheme of the molecules is not unique, but the parametrization of aforementioned parameters with respect to the structures aims to both satisfy physics and chemistry, either by cross-checking titration, solubility and experimental measurements, or verifying that parameters choices make sense from the chemical point of view. The structures of the molecules we model, strongly follow the ones drawn in Figure 1.

## 3. Results and Discussion

### *3.1. Characterization of polymer solutions*

3.1.1. Determination of the overlapping concentration of the polymers

The polyelectrolyte solutions used in this study are relatively concentrated, thus for a better understanding of their behavior, information needs to be collected about their conformational state in solution, which generally impacts decisively on the interactions of the polymers with solid surfaces [13]. The overlapping concentration of both polymers, $c^*$, can be deduced from DLS experiments [61]. This property is important because from a phenomenological point of view, the behavior of polymers differs depending on whether the polymer chains are isolated (concentration, $c$, below $c^*$) or are forming a cross-linked mesh ($c > c^*$) [16]. The determination of $c^*$ was done by the analysis of the changes of the dynamic behavior of the polymer chains in solution, in terms of the changes of the apparent diffusion coefficient ($D^{app}$), as a result of the variation of the polymer concentration in the 0.2 - 30 g/L range. For the sake of example, Figure 2 shows a comparison of the intensity autocorrelation functions for samples of both polymers.

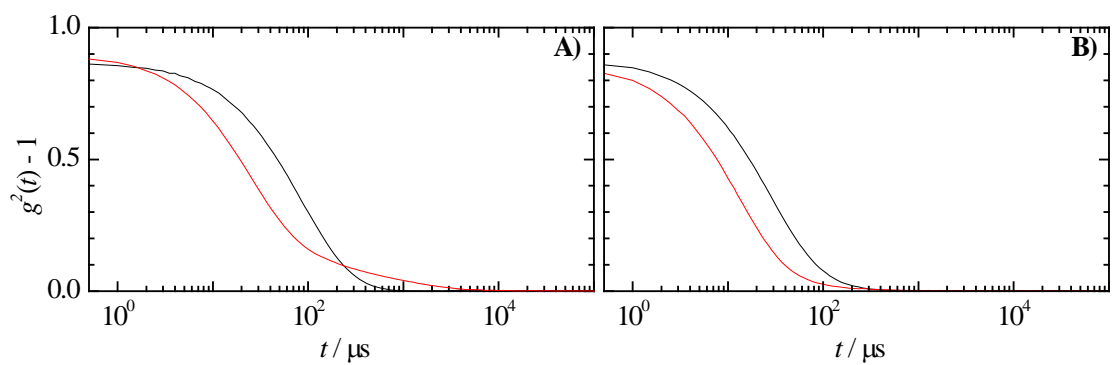

**Figure 2.** Comparison of the normalized intensity autocorrelation functions obtained by DLS experiments for polymer solutions at polymer concentrations below and above c*. (A) Merquat 2003 solutions: (—) 0.2 g/L ($c < c^*$) and (—) 20 g/L ($c > c^*$). (B) PDADMAC solutions: (—) 4 g/L ($c < c^*$) and (—) 15 g/L ($c >$

$c^*$). Note that the polymer solutions have a NaCl concentration of 120 mM and pH fixed at 5.5.

The normalized intensity autocorrelation functions obtained for solutions of both polymers evidence a change on the dynamic behavior of the solutions with the increase of the polymer concentration. Thus, solutions with $c < c^*$ presents a clear single-exponential character, whereas the emergence of a bimodal dynamics was found when $c$ overcomes $c^*$. Figure 3 shows the dependences of the apparent diffusion coefficient, obtained from the analysis of the normalized autocorrelation functions, on the polymer concentration for solutions of both polymers. These dependences provide a method to determine the polymer concentration corresponding to the overlapping threshold, $c^*$.

Despite the complexity associated to the dynamic behavior of polyelectrolyte in solutions, we can rationalize the observed dependence of $D^{app}$ on the polyelectrolyte concentration. For concentrations below $c^*$ we expect for charged Brownian objects, like charged nanoparticles, ionic micelles or polyelectrolytes [62], that the apparent diffusion coefficient increases with concentration due to electrostatic repulsion, this is what it is observed with both polyelectrolytes, although in lower extension for Merquat 2003, that seems to be less charged than PDADMAC and close to theta conditions, at the given ionic strength and temperature. It should be noted that the dependence of the $D^{app}$ on the polymer concentration is exactly the opposite observed for non-charged polymers, in which $D^{app}$ decreases with concentration above $c^*$.

The abrupt change in the slope of the dependence of $D^{app}$ on the polymer concentration, can be associated with the transition to the semi-diluted region and the corresponding $c^*$ can be estimated. For both polyelectrolytes, above $c^*$ a second slow relaxation mode emerges in the dynamic behavior of the solutions. It is worth to mention that even the complete overlapping regime occurs much above $c^*$, and the transition region is very complex in terms of the dynamic behavior it is known that thermodynamic interactions (in the studied system electrostatic interactions) prevail over hydrodynamic interactions, that become screened. On the bases of the experimental results, it may be assumed that in the studied polymer electrostatic interactions also dominate the dynamic behavior in the $c>c^*$ region resulting in a further increase of $D^{app}$ with concentration. However, it should be stressed that this does not mean that the hydrodynamic radius is changing. The hydrodynamic radius can only be obtained from $D^{app}$ (Stokes-Einstein equation) in the diluted region by extrapolation to c→0. The increase in $D^{app}$ with concentration, both in the diluted and semi-diluted region, it is mainly related with the effect of the electrostatic repulsion, and hence only an appropriate and complex analysis, which is out of the scope of this paper, considering electrostatic and hydrodynamic interactions could properly account for the observed values of $D^{app}$.

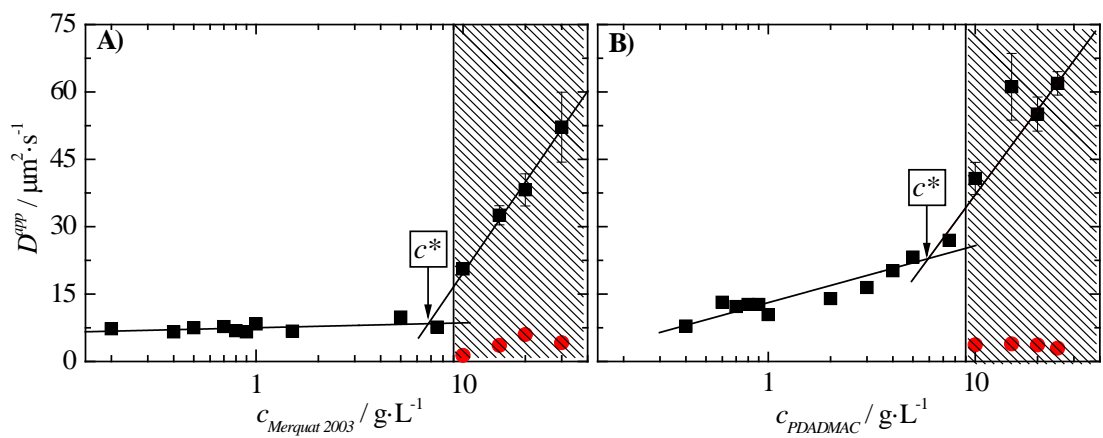

**Figure 3.** Polymer concentration dependence of the apparent diffusion coefficient $D^{app}$ for Merquat 2003 (A) and PDADMAC (B) solutions. The different symbols indicate the two dynamics modes appearing in polymer solutions: (■) Brownian diffusion and (●) slow mode. The solid lines are guides for the eyes, and the crossing point between the two lines can serve as approach to the estimation of $c^*$. The shadowed region represents the concentration range where polymer chains form cross-linked mesh. Note that the polymer solutions have a NaCl concentration of 120 mM and pH fixed at 5.5.

The second dynamic process (slow mode) appearing in time-scales above that corresponding to the Brownian diffusion of the polymer chains for the highest polymer concentrations is probably associated with the elastic behavior of the polymer network, remaining rather independent on the polymer concentration. Therefore, the results allows defining the overlapping concentration, $c^*$, as the point in which the crossover between the two dynamics regimes occurs: (i) Brownian diffusion with electrostatic interactions at low polymer concentrations ($c < c^*$), and (ii) Brownian diffusion with electrostatic interactions + elastic relaxation of the polymer network at high polymer concentrations ($c > c^*$). This approach allows estimating the overlapping concentration at concentrations about 6.7 g/L and 6 g/L for Merquat 2003 and PDADMAC, respectively. The appearance of the slow mode is explained as resulting from the percolation process that usually occurs for concentrations above $c^*$. Thus, it should be expected that even though at $c^*$ the overlapping of the polymer chains starts, the formation of a true cross-linked polymer mesh does not occur up to higher polymer concentration values (around 9 g/L for both polymers). This justify the emergence of the slow mode for polymer concentrations well above of the overlapping one [63].

The similarity of the $c^*$ values for Merquat 2003 and PDADMAC is an unexpected result because the higher molecular weight of Merquat 2003 would suggest that its $c^*$ should appear at much lower concentration than that of PDADMAC. However, the higher charge density of PDADMAC chains results in a strong repulsive interaction between the monomers, and hence, PDADMAC chains adopt a rod-like conformation in solution which increases the gyration radius of the polymer chains, favoring their overlapping. On the other side, the lower charge density of the Merquat 2003 chains results in a more coiled conformation of the polymer in solution, which leads to an increase of the $c^*$ value compared to that expected for a fully charged polyelectrolyte with the same length of Merquat 2003. Therefore, the effect of the larger size of Merquat 2003 is compensated by its lower charge density in relation to PDADMAC, leading to close values of $c^*$ for both polymers.

The extrapolation of the apparent diffusion coefficients to infinite dilution in the diluted regime (below the overlapping concentration) provides information about the infinite dilution diffusion coefficient $D_0$, with $D_0$ (PDADMAC) = $1.4 D_0$ (Merquat 2003). This indicates that an isolated PDADMAC chain diffuses faster than a Merquat 2003 one, which seems to be reasonable considering the lower molecular weight, and corresponding size of the former polymer. It should be noted that a comprehensive discussion on the correlations between the diffusion coefficient and the molecular size, requires considering the differences on the shape of the polymer chains. This is especially important when anisotropic particles, such as polymer chains, are considered because the shape of the chains may influence their dynamic behavior.

3.1.2. Characterization of the Merquat 2003 and PDADMAC solutions

Polymer solutions are characterized by their high viscosities, that are correlated to the nature of the polymer, the concentration of the solutions, the average molecular weight of its chains and as matter of fact to the packing degree of the chains in the solution. Figure 4A shows the polymer concentration dependences of the reduced viscosity $\eta_{red}$, calculated from the solution viscosity $\eta$ (Figure 4B), for Merquat 2003 solutions prepared both in pure water and in NaCl solutions of 120 mM concentration. The results show an increases $\eta_{red}$ with the concentration independently of the salt concentration, with the hydrodynamic volume of the chain, obtained by an extrapolation to infinite dilution ($c_{polymer} \rightarrow 0$), in absence of salt being almost three times higher than the hydrodynamic volume of the chains in presence of salt. The higher values of $\eta_{red}$ and the hydrodynamic volume found in absence of salt are explained considering that the electrostatic repulsion between the monomers leads to more extended conformation of the polymer chains in absence of salt, making the flow of the fluid through the solution difficult. On the other hand, the addition of salt favors the flow of the fluid and decreases the viscosity as a result of the increase of the free volume in the solution associated with the presence of polymer chains in a more coiled conformation.

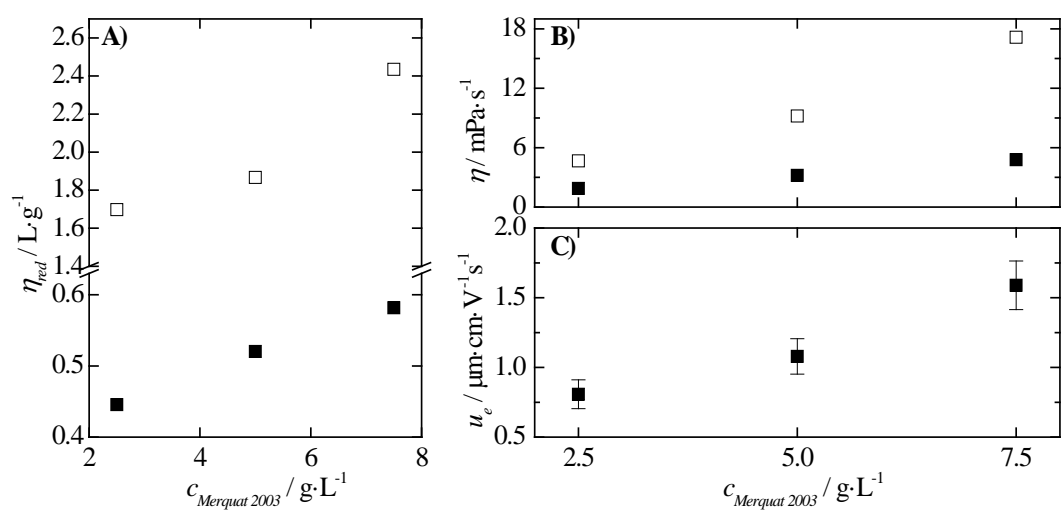

**Figure 4.** Concentration dependences of the reduced viscosity $\eta_{red}$ (A), viscosity $\eta$ (B) and electrophoretic mobility $u_e$ (C) for Merquat 2003. Solutions with a NaCl concentration of 120 mM (■) and Solutions without NaCl (□). The pH of the solutions was fixed at 5.5.

Figure 4C shows the concentration dependence of the electrophoretic mobility of Merquat 2003 in presence of NaCl (*Note: the low conductivity of salt-free solutions and their high viscosity make it difficult to obtain electrophoretic mobility values with a physical meaning. However, similar concentration dependence and sign should be expected than for those containing salt. Furthermore, the values in absence of salt should be higher than those found*

*in presence of salt due to the lower charge screening*). The dependence of the electrophoretic mobility on polymer concentration can be understood from Equation (5). The increase of the viscosity with the polymer concentration, combined with that of the electrophoretic mobility, suggests an increase of the potential $\zeta$, i.e. the effective charge of the chains in solution. This is explained considering that maintaining the salt concentration constant while the polymer concentration is increased results in an increase of the effective charge density of the chains due to the reduction of the ion condensation, i.e. less salt ions per each polymer chain.

We characterized the PDADMAC solutions following the same procedure above described for Merquat 2003 (see Figure 5). The reduced viscosities (see Figure 5A) for PDADMAC solutions in absence of salt appear higher than those containing NaCl, which can be explained as for Merquat 2003 solutions. However, the concentration dependences of $\eta_{red}$ for PDADMAC and Merquat 2003 were very different, showing a clear dependence of the presence of salt. The reduced viscosity of solutions in absence of salt decreases with the increase of the polymer concentration, even though as expected $\eta$ increases with polymer concentration (see Figure 5B). This is because of the contraction of the PDADMAC chain when the solution concentration decreases. This is explained by considering the screening of the inter-monomer electrostatic repulsive interactions due to the increase of the ionic strength associated with the increase of the polymer concentration [64]. On the other hand, $\eta_{red}$ remains almost constant with the polymer concentration for solutions containing NaCl. This suggests that the size of the chains remains almost invariable with the change of polymer concentration in solutions with NaCl. It is worth mentioning that the viscosity of PDADMAC is significantly lower than the viscosity found for Merquat 2003 (see Figure 5B). This may be rationalized considering the huge difference in the average molecular weights ($M_w$) between both polymers ($M_w$ Merquat 2003 $\approx 6 M_w$ PDADMAC). Hence, taking into account that the size of the polymer chains is one of the most critical factors governing the viscosity of the polymer solutions, the lower viscosity of PDADMAC solutions compared to those of the Merquat 2003 is justified. Notice that for PDADMAC and Merquat 2003 with the same molecular weights, it would be expected a higher viscosity for the solutions of the former polymer due to its more extended conformation.

Figure 5.C displays a comparison of the electrophoretic mobilities obtained for both polymers, and shows that they are not significantly different. This can be explained considering again Equation (5). Thus, the huge difference of the molecular weights of the polymers makes it reasonable to expect that Merquat 2003 can present a $\zeta$ potential higher than PDADMAC.

Furthermore, the electrophoretic mobility, and consequently the ζ potential, of PDADMAC also increase with the concentration, which is explained again by the reduction of the charge screening as the polymer concentration increases.

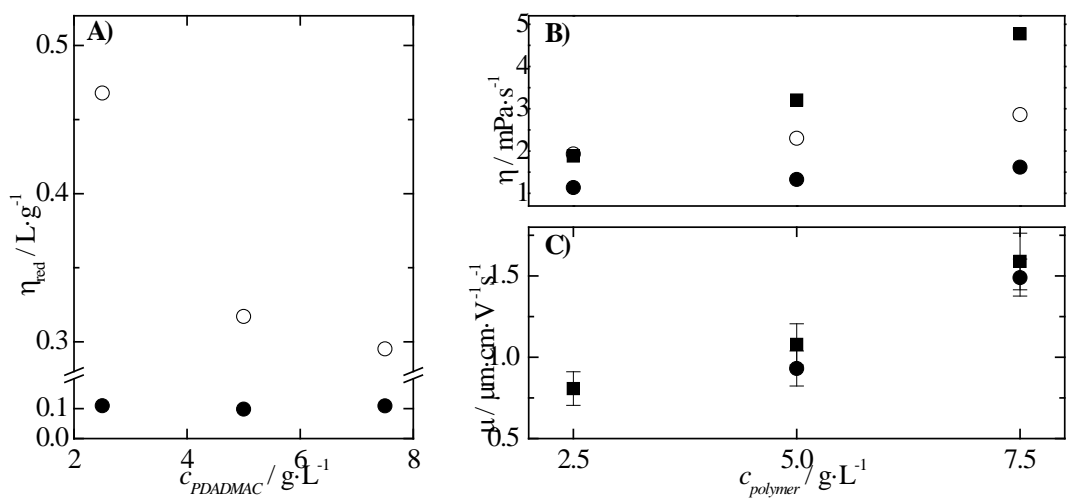

**Figure 5.** Concentration dependences of the reduced viscosity $\eta_{red}$ (A), viscosity $\eta$ (B) and electrophoretic mobility $u_e$ (C) for PDADMAC. Solutions with a NaCl concentration of 120 mM (●) and Solutions without NaCl (○). In panels (B) and (C) the data for solutions of Merquat 2003 with a NaCl concentration of 120 mM are reported for the sake of comparison (■). The pH of the solutions was fixed at 5.5.

3.1.3. Characterization of solutions of PDADMAC:Merquat 2003 mixtures

Above we have focused the discussion on the analysis of the physico-chemical behavior of solutions containing only PDADMAC or Merquat 2003. In the following, we will discuss the results obtained for the characterization of mixtures of both polymers, with total concentrations of 5 g/L and 7.5 g/L. Considering the similar value of $c^*$ found for both polymers, it would be expected that in the range of concentrations explored, the chains of both polymers appear isolated. Therefore, no interference in the interactions between the polymers associated with the formation of cross-linked polymer are expected. Figure 6A shows the dependence of the viscosity on the concentration of each polymer in the mixture (mixtures with a concentration total of polymer of 5 g/L). The results show that the viscosity of the solutions decreases following a pseudo-linear dependence with the increase of the PDADMAC content in the mixture, the opposite occurs with the increase of the Merquat 2003 concentration. This suggests that the behavior of the mixtures is governed by the main component of the mixture. Furthermore, the viscosity of mixtures containing both PDADMAC and Merquat 2003 decreases with the increase of the ionic strength in a similar way to that previously observed for solutions containing only one of the polymer. This may be explained again considering an enhanced flow of the fluid due to a conformational change of the polymer chains as a result of the charge shielding through the condensation of the counterions.

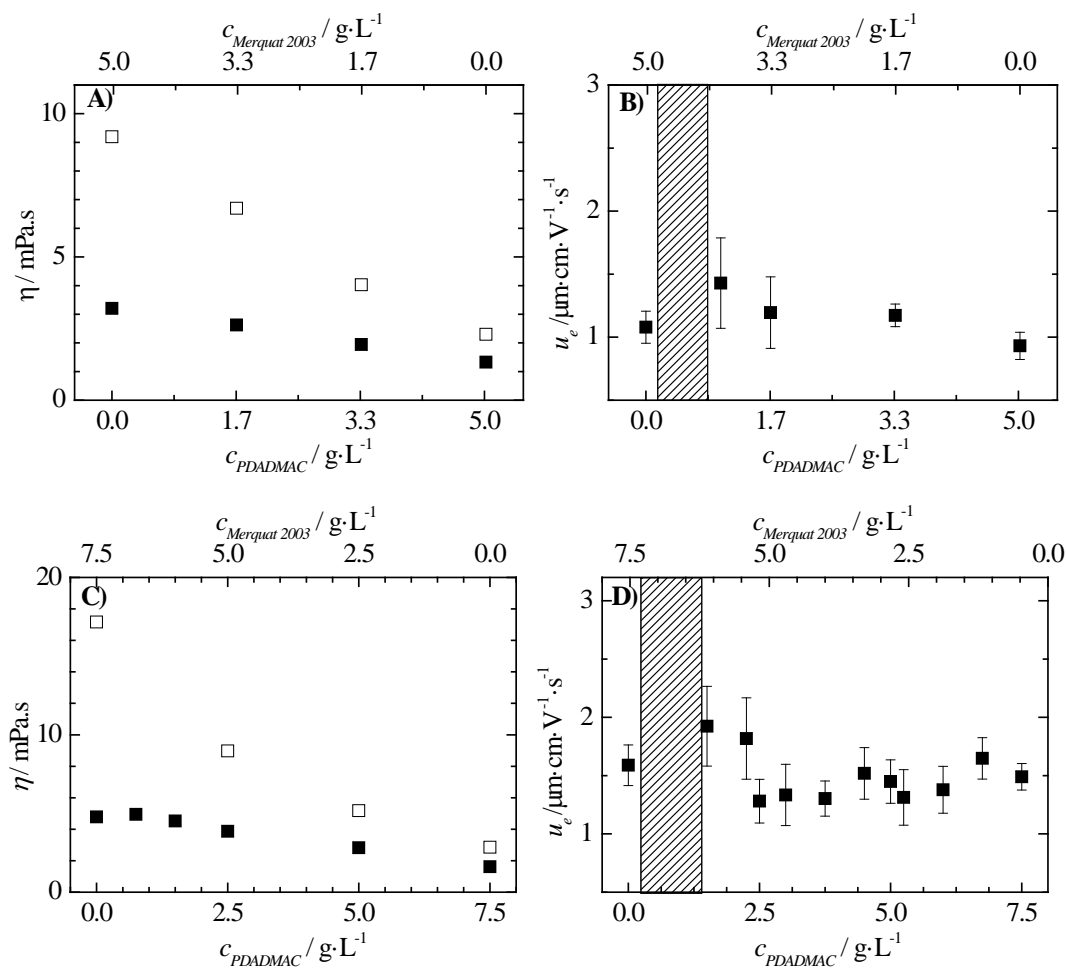

**Figure 6.** (A) Dependences of the viscosity $\eta$ on the concentration of each polymer for PDADMAC:Merquat 2003 mixtures with a total polymer concentration of 5 g/L. (B) Dependence of the electrophoretic mobility $u_e$ on the concentration of each polymer for PDADMAC:Merquat 2003 mixtures with a total polymer concentration of 5 g/L. (C) Dependences of the viscosity $\eta$ on the concentration of each polymer for PDADMAC:Merquat 2003 mixtures with a total polymer concentration of 7.5 g/L. (D) Dependence of the electrophoretic mobility $u_e$ on the concentration of each polymer for PDADMAC:Merquat 2003 mixtures with a total polymer concentration of 7.5 g/L. Solutions with a NaCl concentration of 120 mM (■) and solutions without

NaCl (□). The shadowed region in panels (B) and (D) represents a region of high turbidity. The pH of the solutions was fixed at 5.5.

The interaction of PDADMAC and Merquat 2003 in solution can be evaluated from the dependence of the electrophoretic mobility on the concentration of both polymers shown in Figure 6B. The results show that the electrophoretic mobility remains almost constant and positive, independently on the weight fraction of each polymer in the mixture. This means that the effective charge of the mixtures decreases with the PDADMAC concentration down to reach the value corresponding to that of the PDADMAC chains. This suggests that despite the positive net charge of both polymers, the mixing of PDADMAC and Merquat 2003 results in the formation of inter-polymer complexes in the aqueous solution due to the interaction of the positively charged monomers of PDADMAC with those negatively charged of the copolymer, with their effective charge depending on the main component of the mixture.

Further information about the interaction of both polymer in solutions were obtained from the region of instability appearing for mixtures with a relatively low PDADMAC concentration (shadowed region in Figure 6B). The mixtures within the composition range associated with the instability region present a high turbidity (*Note that electrophoretic mobility measurements or DLS one cannot be performed within turbid regions*), even though no signature of sedimentation were found in the dispersion. This confirms that the mixture of PDADMAC and Merquat results in the formation of inter-polymer complexes. However, the turbidity cannot be explained only assuming the formation of inter-polymer complexes. It is commonly accepted that mixtures of oppositely charged polyelectrolytes undergo phase separation when stoichiometric complexes (same number of positive and negative charges are present in the solution) [65-68]. This is not the case in PDADMAC: Merquat 2003 mixtures within the explored compositional range, where the formation of positively charged complexes is expected for all the studied mixtures. Thus, the emergence of an instability region suggests the formation of kinetically trapped aggregates similar to those formed in polyelectrolyte-surfactant mixtures [69-73]. This occurs as a result of a local excess of one of the polymers during the mixing process, which leads to the formation of aggregates presenting a quasi-neutral inner core, even if their external region remains charged. This surface charge provides them with colloidal stability, hindering its sedimentation. It should be expected that the formation of kinetically trapped aggregates strongly impacts the deposition pathway as was demonstrated for such type of aggregates in polyelectrolyte-surfactant mixtures [5, 7].

Figures 6C and 6D show that the increase of the total polymer concentration until 7.5 g/L does not modify the qualitative dependences of the viscosity and the electrophoretic mobility on the polymer concentration, and the only difference is the expected increase of both parameters as a result of the increase of the total polymer concentration in the solution. Furthermore, the increase of the total polymer concentration results in a broadening of the region of instability of the mixtures.

Additional insights on the behavior of the polymer mixtures were obtained by studying their dynamic behavior in solutions using DLS. For the sake of example, Figure 7A shows some selected normalized intensity autocorrelation functions corresponding to mixtures with different weight ratios of the two polymers and a fixed polymer concentration of 5 g/L.

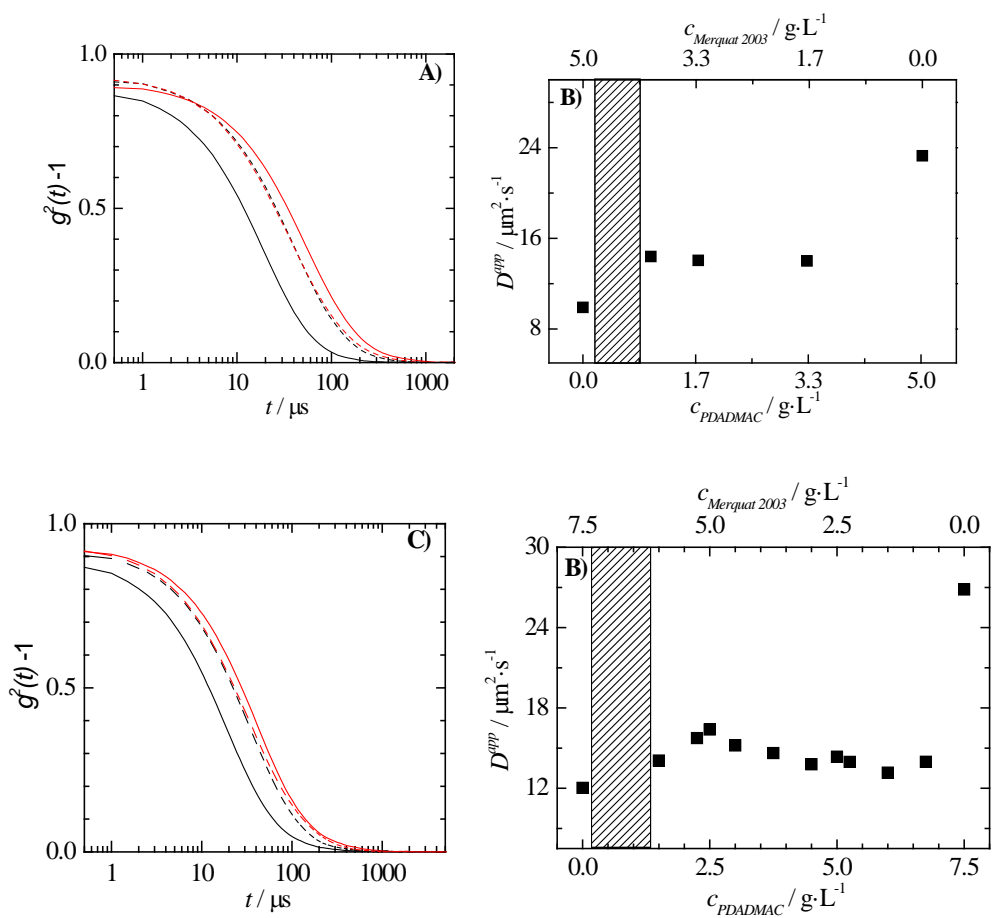

**Figure 7.** (A) Normalized intensity autocorrelation functions obtained by DLS experiments for solutions with total polymer concentration of 5 g/L: (—) and (—) Merquat 2003 and PDADMAC solutions, respectively, (---) PDADMAC:Merquat 2003 mixture (weight ratio 1:2) and (---) PDADMAC:Merquat 2003 mixture (weight ratio 2:1). (b) Dependence of the apparent diffusion coefficient $D^{app}$ on the concentration of each polymer for mixtures of both polymers with a total polymer concentration of 5 g/L. (C) Normalized intensity autocorrelation functions obtained by DLS experiments for solutions with a total polymer concentration of 7.5 g/L: (—) and (—) Merquat 2003 and PDADMAC solutions, respectively; (---) PDADMAC:Merquat 2003 mixure (weight ratio 1:2) and (---) PDADMAC:Merquat 2003 (weight ratio 2:1). (D) Dependence of the apparent diffusion coefficient $D^{app}$ on the concentration of each polymer for mixtures of both polymers with a total polymer concentration of 7.5 g/L. The shadowed region in panels (B) and (D) represents a region of high turbidity. The pH of the solutions was fixed at 5.5 and the NaCl concentration was 120 mM.

The results show that all the normalized intensity autocorrelation functions present a clear single-exponential character, with their characteristic relaxation presenting an intermediate behavior between the pure polymers. The absence of two relaxation processes with different time-scales allows one to disregard the existence of two species with different sizes diffusing simultaneously in the solution, which should be considered a clear confirmation of the formation of inter-polymer complexes in the solution. The apparent diffusion coefficients of the mixtures (Figure 7B), obtained from the analysis of the normalized autocorrelation functions of intensities, confirm the existence of an intermediate dynamic behavior between those corresponding to the pure polymers, with the complexes showing a dynamic behavior closer to that corresponding to pure Merquat 2003. This suggests that the bigger size of Merquat 2003 defines the size of the formed inter-polymer complexes, and hence their dynamic behavior remains reminiscent of that found in Merquat 2003 solutions. It is worth mentioning that the increase of the total polymer concentration to 7.5 g/L (see Figure 7C and 7D) does not modify significantly the above discussed picture, and only a slight increase of the value of the $D^{app}$ was found with the polymer concentration.

*3.2. Adsorption of polymer mixtures onto negatively charged solid surfaces*

3.2.1. Thickness and water content of layers PDADMAC:Merquat 2003 mixtures

We concluded above that the adsorption layers generally mirror the interactions and aggregation phenomena occurring in solution. In the following, the adsorption of PDADMAC:Merquat 2003 mixtures onto negatively charged surfaces will be discussed. It is expected that the positive charge of the inter-polymer complexes, evidenced on the positive values of the electrophoretic mobility, have a significant impact on the adsorption. However, the formation of kinetically trapped aggregates should contribute further to the adsorption within the corresponding concentration range where they are formed. Figure 8 shows the dependences of the $h_{ac}$ and $h_{op}$ on the concentration of the polymers for the adsorption of layers of polymer mixtures deposited from solution with a fixed NaCl concentration of 120 mM. It should be noted that the discussed values for the thickness correspond to those obtained upon rinsing the layer with an aqueous solution at pH=5.5 of NaCl with concentration of 120. This rinsing process does not modify significantly the adsorbed amount, which allows one to consider an irreversible adsorption for the polymers. Furthermore, it is worth mentioning that the viscoelastic character of the polymer layers, evidenced in two characteristics of the results obtained from QCM-D ((i) the absence of overlapping upon the polymer adsorption between the frequency shifts corresponding to the different overtones of the quartz sensors, and (ii) the relatively high values of the dissipation factor), made necessary to use a model considering the viscoelastic properties of the layers for obtaining the $h_{ac}$ values.

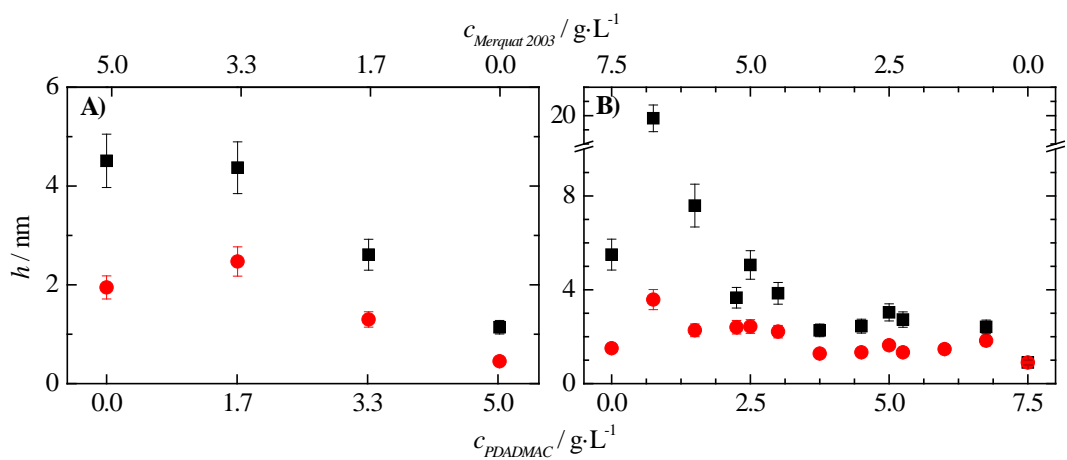

**Figure 8.** Dependences of $h_{ac}$ (■) and $h_{op}$ (●) on the concentration of each polymer for layers of different PDADMAC:Merquat 2003 mixtures as were deposited from solution with two different total polymer concentrations: 5 g/L (A) and 7.5 g/L (B). The pH of the solutions was fixed at 5.5 and the NaCl concentration was 120 mM. The layer thickness values correspond to that obtained upon rinsing the films with an aqueous solution of pH=5.5 and NaCl concentration of 120 mM.

The results show that $h_{ac} > h_{op}$ over the whole total polymer concentration range, and the composition of the mixture. This can be explained considering the different sensitivity of the techniques to the associated water as was stated above. The detailed analysis of the adsorbed amount points out that Merquat 2003 leads to thicker layers than those of PDADMAC due to the chemical composition of both polymers. The existence of neutral, cationic and anionic monomers in Merquat 2003 results in the formation of fuzzy layers, with many neutral and anionic segments distorting the layer structure. This forces the polymer chains to an adsorbed conformation where many loops and tails appear protruding to the aqueous solution. This is not possible in PDADMAC layers, in which the electrostatic repulsion between the charged monomers flattens the polyelectrolyte chains deposited onto the surface, and hence limiting the number of polymer segments protruding to the solution. Similar differences were reported in our previous publication for the deposition of PDADMAC and different copolymers containing blocks with different charge onto negatively charged surfaces [2]. Notice that the flat adsorbed conformation implies an entropy penalty to the adsorption process.

Moreover, the dependences of the adsorbed amounts on the polymer concentrations suggest that the behavior of the polymers and their mixtures in solution markedly affects the adsorption process. Thus, the thickness of the layers appears, within almost the entire range of explored composition, as intermediate between those corresponding to the layers of the single polymer. However, this situation changes when the adsorption of layers occurs for mixtures belonging to the compositional range in which an instability of the solutions of the mixtures appears. The thickness of the layers evidences a sharp increase with the onset of the instability region, and then the thickness decreases as the PDADMAC concentration in the mixture is increased. This scenario is that found for the adsorption of polyelectrolyte-surfactant mixtures with compositions belonging to their phase separation region [12, 70]. The increase of the thickness within the instability region can be attributed to an enhanced deposition due to the sedimentation of the phase-separated kinetically trapped aggregates. Therefore, the adsorption within these layers can be assumed to occur through two concurrent process: (i) electrostatically driven deposition of the inter-polymer complexes, and (ii) gravitational transport of the kinetically trapped aggregates [74, 75]. For the adsorption of pure polymer and mixtures corresponding to other compositional regions, only the first process is relevant.

The comparison of the layer thicknesses obtained after deposition from solutions with different polymer concentrations provides evidence that the increase of the polymer concentration results in the increase of the average thickness of the layers due to the competition of the polymer chains for the adsorption onto the surface. This leads to the formation of layers with an increasing number of loops and tail protruding to the solution.

Further insights on the adsorption layers can be obtained from the evaluation of the fraction of water associated with the adsorbed layers using the Equation (9). Figure 9 shows the polymer concentration dependences of the water content $x_w$ for PDADMAC:Merquat 2003 mixtures.

The water content of polymer layers remains relatively constant with the composition of the mixture, at values about 50% of the total weight fractionof the layer for films obtained from solutions with a total polymer concentration of 5 g/L. However, the situation changes significantly when the total polymer concentration increases up to a value of 7.5 g/L, for which the water content decreases with the increase of the PDADMAC concentration. This is explained considering that the increase of PDADMAC in the mixtures results in the adsorption of the polymer or inter-polymer complexes in a collapse conformation to minimize the electrostatic repulsions. On the other hand, the existence of a high concentration of loops and tails when Merquat 2003 is the main component of the mixtures results in an increase of the water content of the layers as a result of the trapping of the water molecules within the fuzzy region of the layers. The increase of the thickness and water content of the layers with the addition of increasing amounts of Merquat 2003 to the mixtures agrees with the scenario proposed by Caetano et al. [76] on the bases of Monte Carlo simulations. They explored the adsorption of polyampholytes containing a random distribution of their monomers along the polymer chain (i.e. similarly to that what happens when Merquat 2003 is considered), and found that such polymers adsorb onto charged surfaces forming layers with fuzzy structure, i.e. containing any loops and tails appears protruding to the solution. This results in the increase of the layer width, i.e. the layer thickness. This agrees with the theoretical model proposed by Cherstvy and Winkler [77], which assumes that the equilibrium adsorption of polyelectrolytes is shifted towards the formation of desorbed states; with the exception of those cases in which the adsorption of highly charged polyelectrolytes onto highly charged surfaced is considered. This assumption also supports the increase of the thickness and water content of the layers with the increase of Merquat 2003 in the mixtures, which may be considered similar to a reduction of the average charge density of the adsorbing polymer chains.

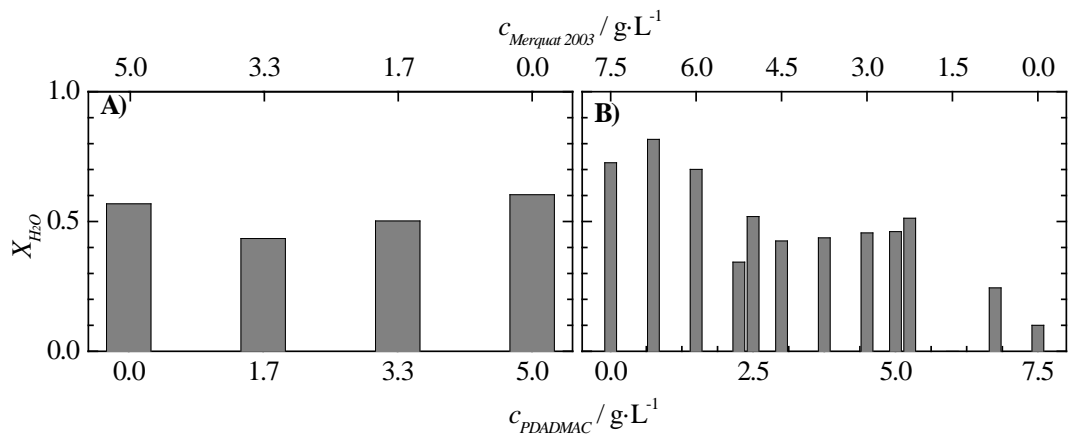

**Figure 9.** Dependences of $x_w$ on the concentration of each polymer for layers of different PDADMAC: Merquat 2003 mixtures as were deposited from solution with two different total polymer concentrations: 5 g/L (A) and 7.5 g/L (B). The pH of the solutions was fixed at 5.5 and the NaCl concentration was 120 mM.

The QCM-D results provide additional insights about the adsorption process in terms of the relationship between the mechanical properties of the film and the amount of adsorbed matter. Figure 10 shows the dependence of the ratio between dissipation factor change and frequency change on concentration of both PDADMAC and Merquat 2003 components [78, 79]. The changes in the dissipation factor are correlated with the ratio between the dissipated and stored energies for the adsorbed layers during the oscillation of the quartz crystal, whereas the frequency changes are related to the amount of polymer and hydration water forming the layer.

The results show a strong increase of the importance of the dissipation factor in relation to the amount of material adsorbed for the mixtures with the lowest PDADMAC. However, considering the decrease of the thickness of the layers as the concentration of PDADMAC in the mixtures increases (see Figure 8), the increase in the $-(\Delta D/\Delta f)$ ratio cannot be associated with the formation of softer layers, but with the formation of thinner layers with the inter-polymer complexes adopting a more collapse conformation onto the surface. Once PDADMAC becomes the main component of the mixture, the $-(\Delta D/\Delta f)$ ratio drops as the concentration of PDADMAC increases. This is explained by a stronger collapse of the adsorbed inter-polymer complexes, which leads to a reduction of the dissipation factor together with the decrease of the thickness.

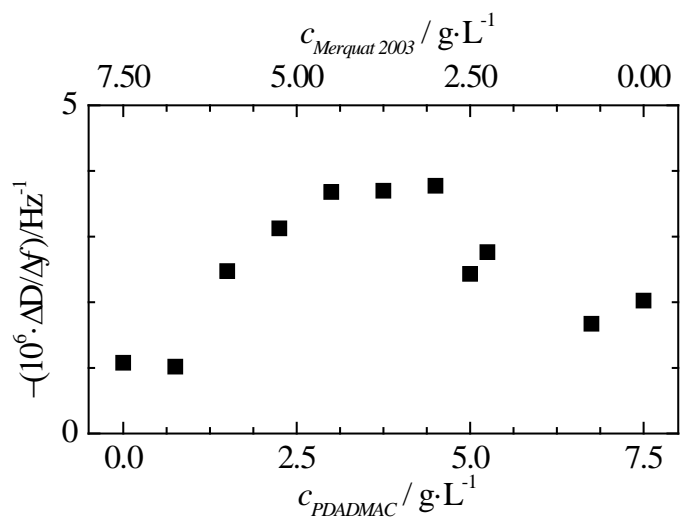

**Figure 10.** Dependences of the ratio $-(\mathit{\Delta D}/\mathit{\Delta f})$ on the concentration of each polymer concentrations for layers of different PDADMAC: Merquat 2003 mixtures as were deposited from solution with a total polymer concentration of 7.5 g/L (B). The pH of the solutions was fixed at 5.5 and the NaCl concentration was 120 mM.

3.2.2. Tribological properties of layers obtained from mixtures of Merquat 2003 and PDADMAC

The control of the tribological properties of a coating is important because the frictional properties of the film allows controlling stiction and lubrication of surfaces, which can impact decisively in the potential applications of the films [80, 81]. In the following, the similarities and differences of the frictional properties obtained by SFA experiments for selected layers of PDADMAC:Merquat 2003 mixtures will be discussed. Figure 11A shows the Normal Force (F) vs. Separation distance (d) curves obtained for two different layers having PDADMAC:Merquat 2003 weight ratios of 1:9 and 2:1.

The results show the existence of a repulsive force upon approaching the surfaces independently of the composition of the layer, with the force increasing steeply for distances low than 10-20 nm, and a finite thickness layer, of a few nm, remaining trapped between surfaces at the highest loads applied. However, the repulsive force is evident from larger distances with the reduction of PDADMAC concentration. This can be understood considering the non-electrostatic origin of the repulsive force. Thus, the increase of the amount of Merquat 2003 results in the formation of thicker layer with a fuzzy structure leads to a situation in which the steric interaction associated with the forced overlapping of the polymer layers deposited onto the opposite surface occurs for larger separation than those found in layer presenting PDADMAC as the main component. Furthermore, hysteresis, without any sign of adhesive forces, was observed upon the separation of the surfaces, with the retraction curve lying below the approach curve (see Figura 11B). Such hysteresis may be probably associated with the finite approach/retraction velocity and out of equilibrium solvent flow in the adsorbed polymer when surfaces were separated. The magnitude obtained of the normal forces obtained for layers PDADMAC:Merquat 2003 mixtures appears in the same range of those reported by Bouchet et al. [60] for PDADMAC layers.

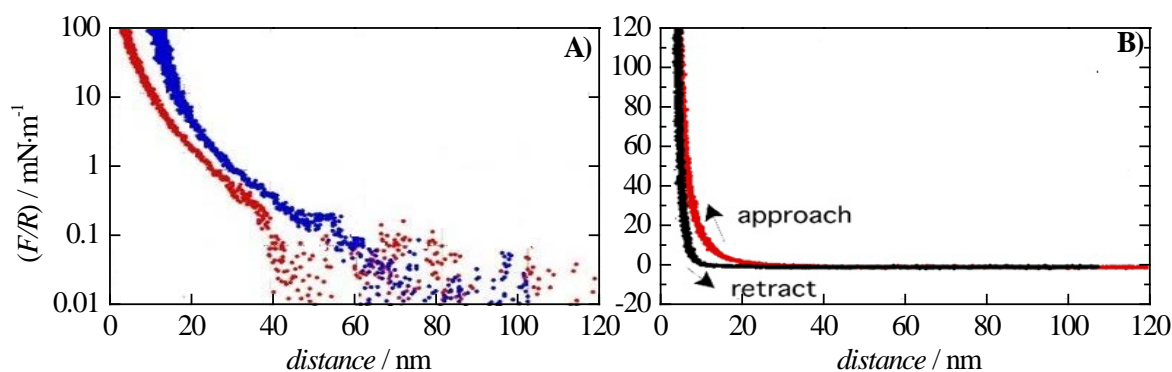

**Figure 11. (A)** Normalized normal load vs inter-surface distance curve (F/R curve) measured upon approaching the surfaces, for layers of PDADMAC:Merquat 2003 mixtures with different weight ratios: 1:9 (■) and 2:1 (■). (B) F/R vs distance for sample with weight ratio 2:1, showing hysteresis upon approach/separation, with no sign of adhesion when separating the surfaces. The pH of the solutions was fixed at 5.5 and the NaCl concentration was 120 mM.

More information about the tribological properties of polymer layers were obtained from the measurements of the friction forces ($T$) under an increasing normal load (F) in the range 500-2300 µN, ($F/R$ ratio from 50 to 230 mN/m), at a fixed shear speed of 10 µm/s (see Figure 12).

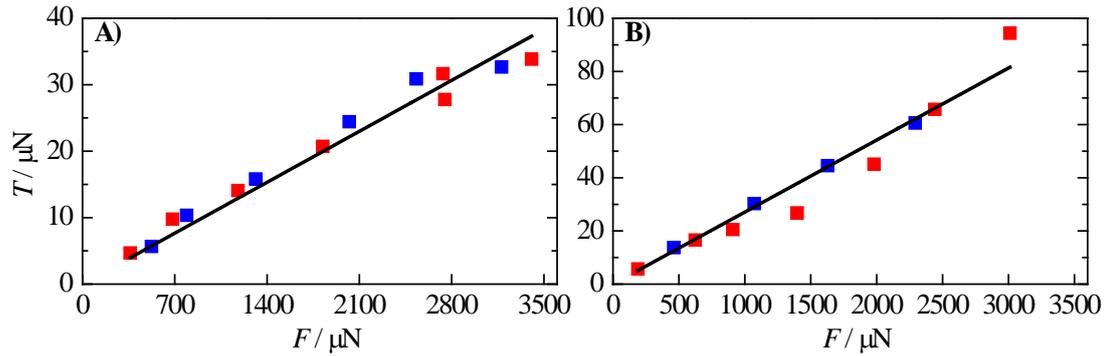

**Figure 12.** Friction force vs normal load at a fixed shear speed of 10 μm/s for layers of PDADMAC:Merquat 2003 mixures with different weight ratios: 1:9 (A) and 2:1 (B). In both panels: (■) increasing normal force and (■) decreasing normal force. The solid lines represent the average dependence of the friction force on the normal load, with the slope being the friction coefficient. The pH of the solutions was fixed at 5.5 and the NaCl concentration was 120 mM.

The friction forces were found to be extremely low, with the friction force ($T$) being proportional to the normal load ($F$) independently of the composition of the layers, with the friction coefficient, defined by the slope of the representation ($\mu=T/F$) being remarkably low. Furthermore, the absence of hysteresis in the friction vs. load curves is signature of the absence of wear in the deposited films. The increase of the PDADMAC amount in the layers results in an increase of the friction coefficient from a value of 0.01 for the PDADMAC:Merquat mixtures with weight ratio 1:9 up to a value almost 3-fold higher (0.028) for the mixtures with weight ratio 2:1. This suggests that the inclusion of Merquat 2003 improves the lubrication properties of the films. The dependence of the friction on the shear velocity, $v$, revealed a weak, quasi-logarithmic, increase of friction with the sliding speed when Merquat 2003 is the main component of the layers, with this dependence becomes linear with the increase of the PDADMAC content (see Figure 13).

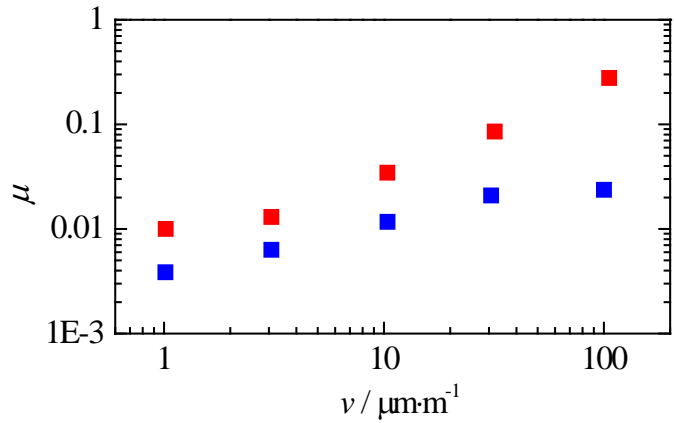

**Figure 13.** Dependence of the Friction coefficient on the shear velocity for layers of PDADMAC:Merquat 2003 mixures with different weight ratios: 1:9 (■) and 2:1 (■) . The pH of the solutions was fixed at 5.5 and the NaCl concentration was 120 mM.

3.2.3 SCF calculations of the adsorption of mixtures of Merquat 2003 and PDADMAC

Figure 14 shows results of SCF calculations of the mixture. In the main panel of Figure 14, the density profiles of both species as a function of the distance, z, perpendicular to the surface, are plotted for different ratio of PDADMAC:Merquat 2003. Such ratios correspond to the symbols (circle, square and star) represented on the adsorption amount curve depicted in the top inset of Figure 14. This master curve informs on the ability of PDADMAC to adsorb when M2003 amount varies and adsorbs on the surface. From this curve, we typically observe an anti-cooperative behavior of both polymers. When Merquat 2003 amount is low, $\theta^S_{M2003} \to 0^+$, adsorbed amount of PDADMAC, $\theta^S_{PDADMAC}$, is high, and vice-versa. Such an anti-cooperative behavior has a drastic effect on the shape of adsorption profiles of the mixtures. In the main panel of Figure 14, we then observe in the first, low $\theta^S_{M2003}$ regime, a thin PDADMAC film of height, $\langle H \rangle$, given in the bottom inset of Figure 14 with the star symbol, whereas Merquat 2003 also forms such kind of film. This result is in agreement with the values obtained experimentally for the adsorption of solution with the highest concentration of PDADMAC (see Figure 8), and finally, both polymers participate to the adsorbed layer and provide its lubrication properties. When the amount of Merquat 2003 slightly increases, the adsorption profiles show the same shape, and film height for PDADMAC slightly decreases whereas a slight increase in Merquat 2003's film height is measured. The trend strongly changes for the highest amount of M2003 considered in the simulations. Figure 14 provides interesting results that rationalize both experimentally measured film thickness shown in Figure 8 and the measured tribological properties of the films in such a regime depicted in Figures 11-13. Of course, SCF calculations do not provide kinetics arguments as only equilibrium situations can be captured within our formalism, however, dynamical effects measured in SFA can find their ground in the structural effects (tails, loops…) captured by SCF. For instance, calculations show a strong increase in film height, $\langle H \rangle$, of M2003 to which adsorption profiles show a fuzzy layer that extend far from the bare surface, which was also discussed in Figure 10.

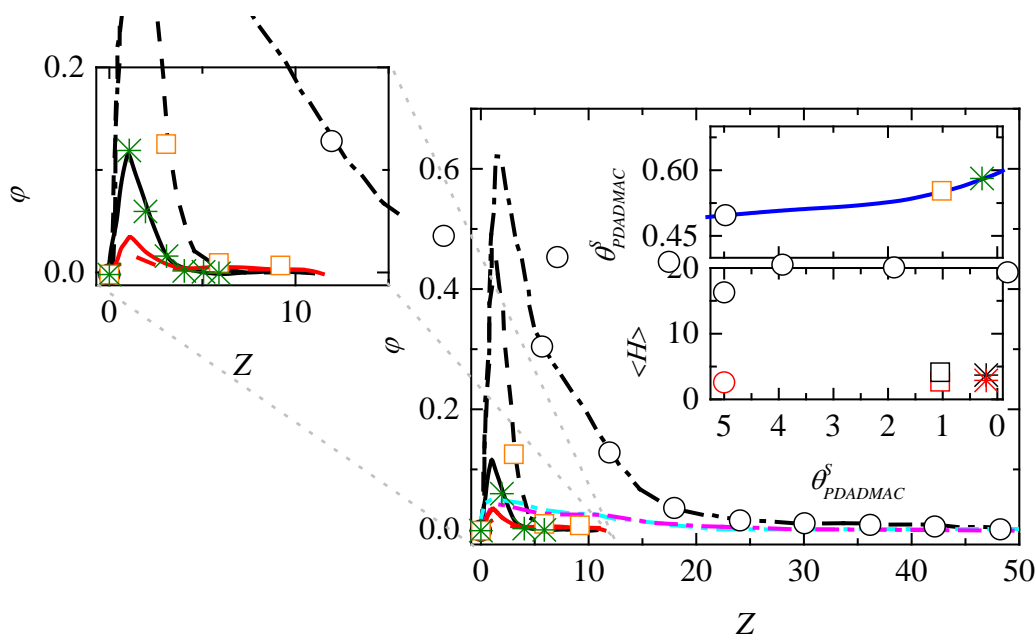

**Figure 14**. Per-species density profiles for the PDADMAC:Merquat 2003 mixture as a function of the distance, z, to the bare surface. The symbols (circle, square and star) represent different ratio of the mixture, following the master curve depicted in the upper inset, where variation of Merquat 2003 amount induces adsorption of PDADMAC as a solution of the SCF problem. For each ratio, the density profiles use full lines (stars), dashed lines (squares) and dashed-dotted lines (circle). In the lower part of the inset, the average layer thickness of each species in the film are also plotted for the corresponding PDADMAC:Merquat 2003 ratio. On the left hand-side of the figure, a zoom closer to the surface is represented.

The profiles calculated in SCF also provide insights on the increase of lubrication of the film when introduction of Merquat 2003 in the mixture is considered, as discussed in Figure 12. From the structural point of view, we strongly believe that molecular species that preferentially adsorb on the negatively charged bare surface are of cationic nature, and most of the tails and loops that provide the fuzzy layer arise from the population of acrylic acid and acrylamide groups which are expected to have good reactivity with water through solvation and hydrogen-bonding. This is confirmed in the profiles calculated from SCF. Indeed, we show in Figure 14 that acrylic and acrylamide (pink and turquoise dashed-dotted lines) groups mostly populate the outermost profiles of the adsorbed films. The formation of the fuzzy layer highly populated by such groups is a possible reason of increased lubrication, the main message being that branched macromolecules indeed may confer increased lubrication properties to the deposed layer, given that branches own chemical groups that may promote preferential interactions with moieties.

## 4. Conclusions

This work combines experiments and theoretical calculations based on the SCF approach for studying the physico-chemical behavior of mixtures formed by different weight fraction of a cationic homopolymer (PDADMAC) and a zwitterionic copolymer (Merquat 2003) in solution and upon adsorption onto negatively charged surfaces. The characterization of the pure polymer solutions has shown that, under the studied conditions, the overlapping concentration of the polymers is similar, appearing for a concentration above the concentration range explored in this study, which has facilitated the analysis of the obtained results. The results have shown that even though both polymers present a completely different behavior both in solution and upon their adsorption onto negatively charged solid surfaces, the adsorption of the mixtures presents an intermediate to those corresponding to the pure polymers. This is the result of the formation of inter-polymer complexes in solution as results of the PDADMAC:Merquat 2003 interest, which present a behavior close to that of the main component of the mixture. This leads to a situation in which the adsorbed layers mirror the composition and characteristics of the complexes formed in the solution, which reflects the impact on the deposition of the phenomena occurring in the solution. Thus, the increase of proportion of Merquat 2003 in relation to that of PDADMAC in the solutions enhances the deposition and results in layers with a fuzzier structure as results of the different nature of the interactions occurring between the monomers of the Merquat 2003 and the negatively charged surface. On the other side, when the PDADMAC content in the mixture exceeds the content of Merquat 2003, the adsorption process is controlled by the PDADMAC, and the adsorption occurs with the inter-polymer complexes collapsing onto the surface, which results in a quick saturation of the surface and a lower hydration of the layers and thinner. The here study contributes to the understanding on the most fundamental physico-chemical correlations between the behavior of polymers in solution and the driving force governing their deposition onto solid surfaces which can a big impact in numerous applications involving the deposition of polymers. The understanding of these aspects becomes really important on the seeking of new bio-based and biodegradable polymers for the substitution of raw materials currently used in consumer products.

**Author Contributions:**

Conceptualization, L.F.P., E.G., F.O., G.S.L. and R.G.R.; methodology, L.F.P., E.G., L.B. and F.L.; software, L.F.P., E.G. and F.L.; validation, E.G., L.B., F.L. F.O., G.S.L. and R.G.R.; formal analysis, L.F.P., E.G., F.L., and L.B.; investigation, L.F.P., E.G., L.B., F.L., D.V., F.O., G.S.L. and R.G.R.; resources, L.B., F.L, F.O., G.S.L. and R.G.R.; data curation, L.F.P., E.G., L.B., F.L., F.O. and R.G.R.; writing—original draft preparation, E.G., LF.P. and F.L.; writing—review and editing, L.F.P., E.G., L.B., F.L. D.V., F.O., G.S.L. and R.G.R.; visualization, L.F.P. and E.G.; supervision, E.G., G.S.L. and R.G.R.; project administration, G.S.L. and R.G.R.; funding acquisition E.G., F.O., G.S.L. and R.G.R.

**Acknowledgments:**

This work was funded by MINECO (Spain) under grants CTQ2016-78895-R and PID2019-106557GB-C21, by Banco Santander-Universidad Complutense grant PR87/19-22513

(Spain) and by E.U. on the framework of the European Innovative Training Network-Marie Sklodowska-Curie Action NanoPaint (grant agreement 955612). Authors also acknowledge the financial support received from L'Orèal (France). The Centro de Espectroscopia y Correlación of the Universidad Complutense de Madrid is acknowledged for its availability in the use of its facilities.

**Conflicts of Interest:** G.S.L. and F.L. are employed by L'Orèal (France). The rest of the authors declare no conflict of interest.